\documentclass[preprint,aps,amsmath,amssymb,showpacs,pre]{revtex4}
\usepackage{graphicx}
\usepackage{amssymb}
\begin{document}
\title{Solution on the Bethe lattice of a hard core athermal gas with
  two kinds of particles} 
\author{Tiago J. Oliveira}
\email{tiago@ufv.br}
\affiliation{Departamento de F\'isica, Universidade Federal de Vi\c
  cosa, 36570-000, Vi\c cosa, MG, Brazil}
\author{J\"urgen F. Stilck}
\email{jstilck@if.uff.br}
\affiliation{Instituto de F\'{\i}sica and National Institute of Science and 
Technology for Complex Systems, Universidade Federal
  Fluminense, Av. Litor\^anea s/n, 24210-346, Niter\'oi, RJ, Brazil}   
\date{\today}
\begin{abstract}
Athermal lattice gases of particles with first neighbor exclusion have
been studied for a long time as simple models exhibiting a fluid-solid
transition. At low concentration the particles occupy randomly both
sublattices, but as the concentration is increased one of the sublattices 
is occupied preferentially. Here we study a mixed lattice gas with
excluded volume interactions only in the grand-canonical formalism
with two kinds of particles: small ones, which occupy a single lattice
site and large ones, which occupy one site and its first
neighbors. We solve the model on a Bethe lattice of arbitrary
coordination number $q$. In the parameter space defined by the
activities of both particles. At low values of the activity of small
particles ($z_1$) we find a continuous transition from the fluid to
the solid phase 
as the activity of large particles ($z_2$) is increased. At higher
values of $z_1$ the transition becomes discontinuous,
both regimes are separated by a tricritical point. The critical line
has a negative slope at $z_1=0$ and displays a minimum before reaching
the tricritical point, so that a reentrant behavior is observed for
constant values of $z_2$ in the region of low density of small
particles. The isobaric curves of the total density of particles as a 
function of $z_1$ (or $z_2$) show a minimum in the fluid phase.
\end{abstract}


\maketitle

\section{Introduction}
\label{intro}
In the beginning of statistical mechanics, the interest was focused
mainly on fluids, and the pioneering work on phase transitions
concentrated on the liquid-gas transition, an example of this kind is
the van der Waals equation of state \cite{vdw73}. In his
phenomenological theory, van der Waals considered the effect of
hard core excluded volume interactions and attractive interactions in
the thermodynamic behavior of the fluid. While attractive
interactions between molecules are essential to produce the liquid-gas
transition, it was later realized that even if only hard core
interactions are considered, interesting effects arise in the
models. Since in this case all allowed microscopic configurations of
the system have the same energy, such models are athermal, and all
thermodynamic effects are of entropic origin. Much is known
about the continuous version of these models, usually called hard
sphere systems. They were studied by a variety of techniques
\cite{hd86}, and a fluid-solid phase transition is found. It is worth
recalling that in the pioneering work on the Monte Carlo simulational
procedure \cite{mrrtt53}, the physical system studied was a fluid of
hard disks. 

The fluid model with excluded volume interactions only may also be
defined on a lattice, so that the positions occupied by
the particles are restricted to sites of a lattice. In the simplest 
version of such
models, the only constraint is that if a site is occupied by one
particle, no other particles may be placed on it. In the
grand-canonical ensemble, this model reduces to the Ising model
without the interaction term, and is therefore trivially solved. No
singularities are found in the thermodynamic functions, as
expected. More interesting results are obtained if the range of
excluded volume interactions is increased. This leads to a variety of
models, and we refer to a recent paper where some of these models were
studied using simulations for a comprehensive survey of the literature
\cite{fal07}. If a particle placed on a site excludes this site and
its first neighbors, indeed a fluid-solid continuous transition is
found. For bipartite lattices such as the hypercubic lattices, at
lower densities, the particles occupy the lattice sites 
randomly, but above a critical density one of the two sublattices is
occupied preferentially by the particles, so that the order parameter
may be defined as:
\begin{equation}
\psi=2\langle |\rho_A-\rho_B| \rangle,
\end{equation}
where $\rho_{A,B}$ is the number of particles in sublattice $A$ or
$B$, respectively, divided by
the number of sites in the lattice, thus assuming a maximum value
equal to $1/2$. This model has a long history, it has been mentioned
in the classical review by Domb \cite{d58}, and has been studied using
the virial 
expansion and the Bethe approximation by Burley shortly after
\cite{b60}. Since then, the thermodynamic behavior of the model has been
investigated using a variety of  
analytical and numerical methods, which indicate a continuous phase
transition in the Ising universality class. On the square lattice,
precise estimates for the critical chemical
potential $\mu_c \approx 1.33401510027774(1)$ and the density of
particles at the transition $\rho_c \approx 0.3677429990410(3)$  were
obtained using transfer matrix and finite size scaling extrapolation
techniques, and the Ising critical exponents were verified with high
precision \cite{gb02}. A recent simulational study of this model on
the square and cubic lattices may be found in \cite{cnd11}. On the
triangular lattice, this model is known as the hard hexagon model, and
was solved exactly by Baxter \cite{b80}. A continuous transition was
found at the critical activity $z_c=\exp(\mu_c)=(11+5\sqrt{5})/5$. The
critical exponents are in the 3-state Potts universality class, as
would be expected considering that in the high density phase the
hexagons occupy preferentially one of the three sublattices.

\begin{figure}
\begin{center}
\includegraphics[width=5.0cm]{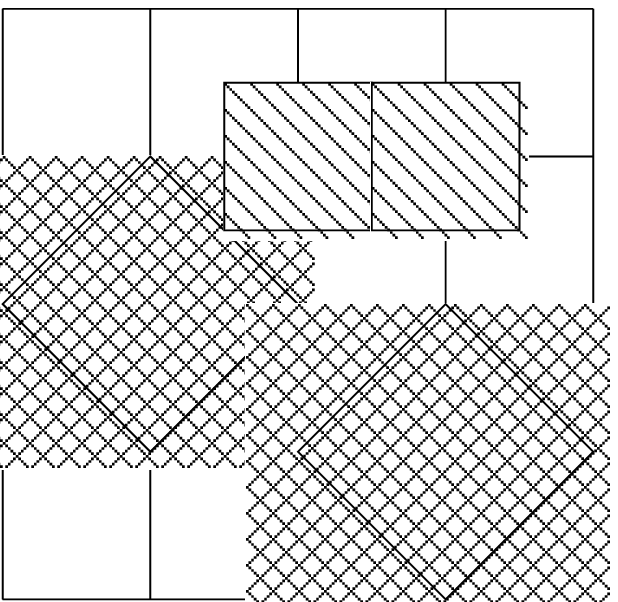}
\caption{Two molecules of type 1 (hatched) and two of type 2
  (cross-hattched) placed on sites of a square lattice. The first may
  be represented as squares of side $a$ and the second as tilted
  squares of side $a\sqrt{2}$, where $a$ is the lattice parameter.}  
\label{f1}
\end{center}
\end{figure}

An interesting generalization of the model is to consider a gas with
both small and large particles, where the small particles occupy a
single site and the large ones the site and its first neighbors. Let
the activity of small (large) particles be $z_1$ ($z_2$). On a square lattice, 
we may represent both particles as squares, as is shown in Fig. \ref{f1}. 
Since a continuous transition occurs when only large particles are present
and no transition is found for small particles only, one may ask what
happens in intermediate situations, with both types particles on the
lattice. This model was studied on the square lattice by Poland
\cite{p84} using high density series expansions in the grand-canonical
formalism, and evidence was found that a tricritical point should be
present in the phase diagram. In this paper we solve the
model on a Bethe lattice of arbitrary coordination number $q>2$ in the
grand-canonical ensemble. As the fugacity of the small particles is
increased starting from zero, the fluid-solid transition remains
continuous up to a certain value, above which it becomes
discontinuous. Therefore, we find a tricritical point in the phase
diagram of the model.

\begin{figure}
\begin{center}
\includegraphics[width=5.0cm]{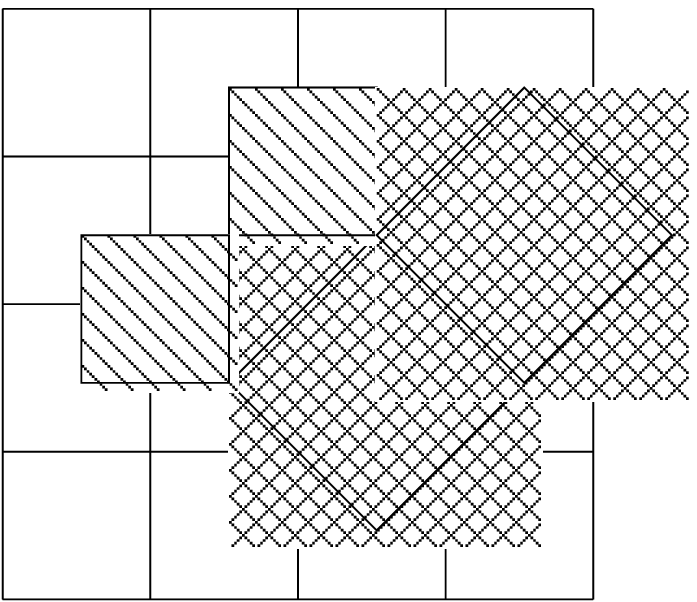}
\caption{Configuration of a model similar that of Fig. \ref{f1}, which 
  was solved in a
  particular case on the square lattice \cite{fl92}. The possible
  locations of  the large particles are dislocated with respect to
  those of the small particles.}  
\label{f2}
\end{center}
\end{figure}

It is worth mentioning that a slight modification of the two-particle
model on the square lattice makes it exactly solvable in a particular
case, by allowing it to be mapped on the Ising model for which the
exact solution is known for zero magnetic field. This particular case
corresponds to $(1+z_1)^2=z_2$ \cite{fl92}. In this lattice gas, the
large particles are placed on the centers of the elementary squares of
the lattice defined by the sites where the small particles are
located, as is illustrated in figure \ref{f2}. Both models are
equivalent in the absence of small particles, but we notice that when
$z_1=0$ we know the exact solution only for $z_2=1$. 

In section \ref{defsol} we define the model and solve it on the Bethe
lattice. The thermodynamic properties of the model are discussed in
section \ref{thermod}, and final discussions and the conclusion may be
found in section \ref{concl}.

\section{Definition of the mixed lattice gas model and its solution on
the Bethe lattice}
\label{defsol}

We study the grand-canonical version of the mixed lattice gas model
defined on a lattice. In this model, two types of particles are
present, and only excluded volume interactions between them are
considered, so that the model is athermal. The particles of type 1
(small) are
such that, when placed on a lattice site, they occupy this single site
only, excluding other particles from it. Particles of
type 2 (large), when placed on a site, occupy its first
neighbor 
sites also. Figure \ref{f1} shows this model on a square lattice and
the generalization to other
hypercubic lattices is straightforward. While no phase transition is
found for the case where only small particles are present, it is well
established that for a pure system of large particles a continuous
phase transition in the Ising universality class occurs
\cite{fal07}. 

Here we will solve the model with both particles present
in the grand-canonical ensemble on the Bethe lattice. The parameters
of the model are the activities of small particles $z_1=\exp(\mu_1)$,
where $\mu_1$ is the chemical potential of a small particle divided by
$k_BT$, and of large
particles $z_2=\exp(\mu_2)$. We proceed defining the model on a Bethe
lattice, 
which is the core of a Cayley tree with general coordination number
$q$. We then consider partial partition functions (ppf's) of
the model 
on subtrees with fixed configurations of the root site, which may be
empty ($0$), occupied by a small particle ($1$) or by a large one
($2$). Considering the operation of connecting $q-1$ subtrees with a
certain number of generations of sites to a new root site, we may
build a subtree with an additional generation and write down recursion
relations for the ppf's. If we call $g_i$ the partial partition
function of a subtree with root site configuration $i=0,1,2$, we have
the following recursion relations for the pff's:
\begin{subequations}
\begin{eqnarray}
g_0'&=&(g_0+g_1+g_2)^{q-1}, \\
g_1'&=&z_1(g_0+g_1)^{q-1}, \\
g_2'&=&z_2g_0^{q-1}.
\end{eqnarray}
\end{subequations}
The prime denotes subtrees with an additional generation. Let us define
ratios of the ppf's as 
$R_i=g_i/g_0$, where now the configuration index $i$
assumes the values 1 and 2. From the recursion relations for the
ppf's, we may obtain the ones for the ratios, which are: 
\begin{subequations}
\begin{eqnarray}
R_1'&=&z_1\frac{(1+R_1)^{q-1}}{(1+R_1+R_2)^{q-1}}, \\
R_2'&=&z_2\frac{1}{(1+R_1+R_2)^{q-1}}.
\end{eqnarray}
\label{rr}
\end{subequations}

The thermodynamic behavior of the model is defined by the values of
the ratios after a large number of iterations of the recursion
relations Eqs. (\ref{rr}). We find that, depending of the values of
the activities, the recursion relations converge either to a fixed
point or to a limit cycle of period 2. For convenience, we may
define the variables $x=1+R_1$ and $y=1+R_1+R_2$. The fixed point
equations in these variables will be:
\begin{subequations}
\begin{eqnarray}
(x-1)y^{q-1}-z_1x^{q-1}&=&0, \\
(y-x)y^{q-1}-z_2&=&0,
\end{eqnarray}
\label{fpe}
\end{subequations}
When the recursion relations converge to a limit cycle, the values of
the ratios in the core of the tree will display a layered structure,
so that the ratios in sites in consecutive generations assume
alternate values. It is convenient, in this case, to define two
sublattices (A and B), whose sites are associated to the two values of
the pair of variables $x,y$. The equations defining this limit cycle
values are:
\begin{subequations}
\begin{eqnarray}
(x_A-1)y_B^{q-1}-z_1x_B^{q-1}&=&0, \\
(y_A-x_A)y_B^{q-1}-z_2&=&0, \\
(x_B-1)y_A^{q-1}-z_1x_A^{q-1}&=&0, \\
(y_B-x_B)y_A^{q-1}-z_2&=&0.
\end{eqnarray}
\label{fpeq}
\end{subequations}
Although we were not able to find general solutions for these
equations, it is not difficult to solve them
numerically. This set of equations may be reduced to finding the roots
of a polynomial in the variable $g=x/y$, given by:
\begin{equation}
h(g)=(1-g)(1+z_1g^{q-1}+z_2f^{q-1})^{q-1}-z_2f=0,
\label{fpeq1}
\end{equation}
where $f=1/y$ is given by:
\begin{equation}
f=g-\frac{z_1}{z_2}(1-g)(1+z_1g^{q-1})^{q-1}.
\end{equation}
The fixed points or limit cycles will correspond to roots of this
polynomial in the range $g \in [0,1]$ with non-negative values for
$f$. Both stable and unstable fixed points will be found. Another
numerical procedure to find the thermodynamic 
properties of the model is to iterate the recursion
relations Eqs. (\ref{rr}) directly, this will lead only to the stable
fixed points. To study the stability of the
fixed points, it is useful to obtain the jacobian of the recursion
relations. The elements of the $2 \times 2$ jacobian
matrix calculated at the fixed point, $j_{i,j}(x,y)=\partial R'_i/\partial
R_j$, are: 
\begin{subequations}
\begin{eqnarray}
j_{1,1}&=&\frac{(q-1)z_1x^{q-2}(y-x)}{y^q},\\
j_{1,2}&=&-\frac{(q-1)z_1x^{q-1}}{y^q},\\
j_{2,1}&=&j_{2,2}=-\frac{(q-1)z_2}{y^q}.
\end{eqnarray}
\label{jac}
\end{subequations}
The stability limit of the fixed point may then be found requiring the
dominant eigenvalue of this to have a unitary modulus. The jacobian
for the limit cycle ${\mathbf J}_2$ will be the product of two
jacobian matrices defined above, calculated at the pair of variables
at the limit cycle, so that $\mathbf{J}_2={\mathbf J}(x_A,y_A) \times
{\mathbf J}(x_B,y_B)$.

The grand-canonical partition function of the model on the Cayley tree
is obtained considering the operation of attaching $q$ subtrees to the
central site of the tree. If the central site is in sublattice A, this
leads to the following expression: 
\begin{equation}
Y_A=g_{0,B}^q(y_B^q+z_1x_B^q+z_2),
\label{pf}
\end{equation}
and a similar expression with the sublattice indexes interchanged is
obtained for sublattice B. It
is easy then to write down expressions for the densities of sites with
small and large particles in the center of the tree. They are:
\begin{subequations}
\begin{eqnarray}
\rho_{1,A}=\frac{z_1x_B^q}{y_B^q+z_1x_B^q+z_2},\\
\rho_{2,A}=\frac{z_2}{y_B^q+z_1x_B^q+z_2},\\
\end{eqnarray}
\label{rho}
\end{subequations}
and the densities on sublattice B are obtained permuting the
sublattice indexes.

The free energy in the core of the Cayley tree, which corresponds to
the Bethe lattice, may be obtained by a generalization of Gujrati's
argument \cite{g95}, which may be found for a particular model in
\cite{cg03}. Here we present a simple derivation generalizing the one
which is found in \cite{osb10}. On the Cayley tree, we admit that the
reduced free energy per site (which corresponds to the grand-canonical
free energy divided by $k_BT$ and the number of sites in the lattice,
which is proportional to its volume) of sites in the $m$'th generation
of the tree 
will be $\phi^{(m)}$. For a tree with $M$ generations, starting to
count at the surface ($m=0$), we may then write the total free energy
as:
\begin{equation}
\Phi^{(M)}=\phi^{(M)}+q\sum_{i=0}^{M-1}(q-1)^i\phi^{(M-i-1)}.
\end{equation}
For a tree with one more generation, the free energy in terms of the
free energies per site will be:
\begin{equation}
\Phi^{(M+1)}=\phi^{(M+1)}+q\sum_{i=0}^{M}(q-1)^i\phi^{(M-i)}.
\end{equation}
Now we notice that:
\begin{equation}
\Phi^{(M+1)}-(q-1)\Phi^{(M)}=\phi^{(M+1)}+q\phi^{(M)}-(q-1)\phi^{(M)}.
\end{equation}
In the thermodynamic limit, we expect the free energies per site to
reproduce the sublattice structure in the core of the tree, and since
sites in consecutive generations belong to different sublattices, we
notice that in both possible cases (A or B sublattice at the central
site) we reach the conclusion:
\begin{equation}
\Phi^{(M+1)}-(q-1)\Phi^{(M)}=\phi_A+\phi_B,
\end{equation}
where the indexes stand for the sublattice. The free energy per site
in the core of the tree is therefore given by:
\begin{equation}
\phi_b=\frac{1}{2}(\phi_A+\phi_B)=-\frac{1}{2}\ln\frac{Y^{(M+1)}}
    {[Y^{(M)}]^{(q-1)}}.
\end{equation}
Using the Eq. (\ref{pf}) and the fixed point equations (\ref{fpe}),
after some algebra, we find the result:
\begin{equation}
\phi_b=-\frac{1}{2}\ln\left[\frac{(y_Ay_B)^{q-1}}
{(y_A-x_A+y_B-x_B+x_Ax_B)^{q-2}}\right].
\label{fe}
\end{equation}
We notice that this expression is invariant under permutation of the
sublattice indexes. For $z_2=0$, where the model
is solved trivially, the Bethe lattice calculation furnishes the exact
solution. The fixed point value in this case is simply $R_1=z_1$, the
reduced free energy per site will be $\phi_b=-\ln(1+z_1)$, and the
density of particles becomes: 
\begin{equation}
\rho_1=\frac{z_1}{1+z_1}=-z_1 \frac{\partial \phi_b}{\partial z_1}.
\end{equation}

\section{Thermodynamic behavior of the model}
\label{thermod}

For simplicity, we will start the study of the thermodynamic
properties of the model in the limit $z_1 \ll 1$, where the density of
small particles is very small. In this region, the fluid-solid
transition is continuous, and the stability limits of both phases are
coincident. To obtain the critical line, we may consider the fixed point
equations (\ref{fpe}) and require the leading eigenvalue of the
jacobian Eq. (\ref{jac}) to be equal to -1, so that:
\begin{equation}
1+j_{1,1}+j_{2,2}+j_{1,1}j_{2,2}-j_{1,2}j_{2,1}=0.
\end{equation}
Now we solve these three equations up to first order in $z_1$,
supposing that $z_2=a+bz_1$, $x=1+b_1z_1$, and $y=a_2+b_2z_1$. We are
thus lead to the following values of the expansion coefficients
defined above: $a=(q-1)^{q-1}/(q-2)^q$, $b=-1$,
$b_1=[(q-2)/(q-1)]^{q-1}$, $a_2=(q-1)/(q-2)$, and $b_2=0$. For the
particular case $z_1=0$, this solution has been obtained a long time ago
\cite{b60}. We also notice that the critical line has a {\em negative}
initial slope. This may be understood
physically  realizing that the presence of few small particles does
lead to an effective entropic attractive interaction between the large
particles, thus favoring their ordering \cite{d10}. At higher values of $z_1$
the slope becomes positive, and finally the transition becomes
discontinuous, the critical line meets the coexistence line at a
tricritical point. 

We may also study the phase diagram in the limit $z_1 \gg 1$. In the
fluid phase, we have the asymptotic fixed point values $x \approx z_1$
and $y \approx z_1$. In the solid phase we have $x_A \approx 1$,
$y_A \approx 1$, $x_B \approx z_1$, and $y_B \approx z_1+z_2$. From
these expressions, we may find the behavior of the coexistence line in
this limit by requiring the free energies, Eq. (\ref{fe}), of both
phases to be equal, and this leads us to the asymptotic behavior $z_2
\approx z_1^2$ for the coexistence line for large values of $z_1$.

\begin{figure}
\begin{center}
\includegraphics[width=8.0cm]{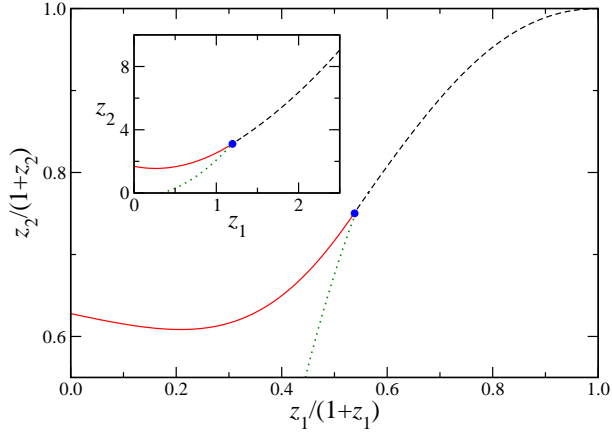}
\caption{(Color on line) Phase diagram of the model in the plane defined by the
  activity fractions $z_1/(1+z_1)$ and $z_2/(1+z_2)$. For a given
  value of $z_1$, the fluid
  phase is stable for lower values of $z_2$, and the solid phase
  becomes stable at higher values. The transition may be continuous or
  discontinuous, both regimes are separated by a tricritical point,
  represented by a circle (blue). The full line (red)
  corresponds to 
  the continuous transition and the coexistence of both phases happens
  at the dashed line. In the inset the same diagram is shown
  with the axes defined by the activities. The dotted line (green) 
  corresponds to the minima of the isobaric curves of the total density
  of particles $\rho=\rho_1+2\rho_2$, as discussed in the text. 
  Bethe lattice with $q=4$.}  
\label{f3}
\end{center}
\end{figure}

\begin{figure}
\begin{center}
\includegraphics[width=8.0cm]{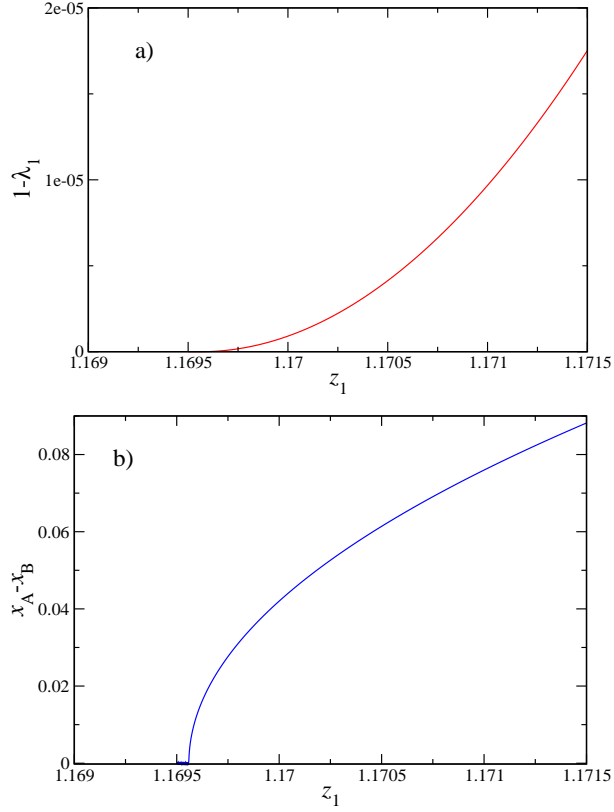}
\caption{a) Behavior of the leading eigenvalue of the
  jacobian of the 
  asymmetric fixed point on the limit of stability line of the
  symmetric fixed point. b) Values of $x_a-x_B$ from the solution of  
  the asymetric fixed
  point equation, calculated on the limit of stability line of the
  symmetric fixed point. Bethe lattice with $q=4$.}
\label{f4}
\end{center}
\end{figure}

The phase diagram in the $(z_1,z_2)$ plane is shown
in figure \ref{f3} for a Bethe lattice with $q=4$, similar diagrams
are found for other values of $q>2$. For $z_1$
larger than the tricritical value, there is an interval of values of
$z_2$ where both fixed points are stable, thus characterizing a
coexistence of both phases, at a value of $z_2$ for which both free
energies are equal. The coexistence line is located between both
stability limit lines, as expected. The numerical determination of the
coexistence line has to be done carefully, particularly close to the
tricritical point, where the range of values of $z_2$ for which both
fixed points are stable becomes very narrow. The precise calculation
of the localization of the tricritical point also demands some
care. It is quite 
easy to calculate the stability limit of the symmetric fixed point,
since the problem may be reduced to finding the solution of an equation in
one variable. Once this line is found precisely, we solve the
asymmetric fixed point equation on it, starting at a value of $z_1$
larger than the tricritical value. As the value of $z_1$ is lowered,
the largest eigenvalue of the jacobian $\lambda_1$ increases and the
values of the variables $x_A$ and $x_B$ which solve the fixed point
equations (\ref{fpeq}) 
become closer, as do the variables $y_A$ and $y_B$. At the fixed
point, the dominant eigenvalue is unitary and $x_A=x_B$. This is
illustrated in figures \ref{f4}, thus leading to an
estimated position of the tricritical point. Table (\ref{t1}) presents
the locations of the tricritical point for several values of the
coordination number $q$, as well as the values of the densities at
this point. An alternative procedure for determining the
location of the tricritical point will be presented below. In all
cases we studied, the critical value of $z_2$ for $z_1=0$,
$z_{2,c}=a$ given above, is smaller 
than $z_2$ at the tricritical point, therefore for $z_2$ below
$z_{2,c}$ and above the minimum critical value of $z_2$ the solution
displays a reentrant behavior as $z_1$ is increased, 
starting in the fluid phase, then getting into the solid phase and
finally ending in the fluid phase again, with two continuous
transitions.  

\begin{figure}
\begin{center}
\includegraphics[width=8.0cm]{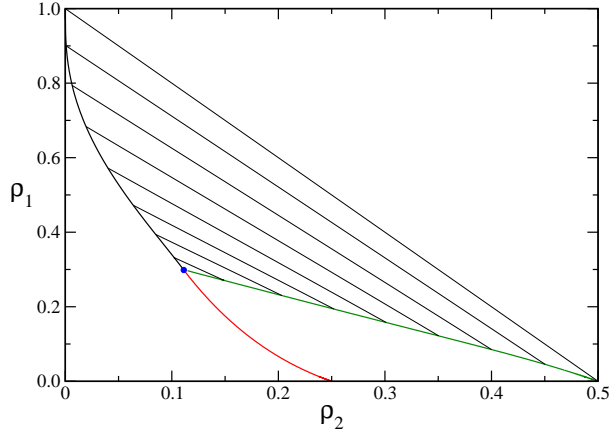}
\caption{(Color on line) Phase diagram in the plane defined by the densities of
  particles $\rho_1$ and $\rho_2$. The critical line (red) ends
  at the tricritical point represented by a circle (blue). The
  densities of the coexisting fluid (black) and solid (green) phases 
  with the same free energies in the two-phase region
  are connected by tielines. Bethe lattice with $q=4$.} 
\label{f5}
\end{center}
\end{figure}

The phase diagram in the plane defined by the densities of small and
large particles is presented in figure \ref{f5}. Again we may obtain
the asymptotic behavior in the limits $z_1 \ll 1$ and $z_1 \gg 1$
using the results for these limits presented above. In
the limit of low density of small particles, we find that the densities
are given by $\rho_1 
\approx (q-2)^qz_1/[q(q-1)^{q-1}]$ and $\rho_2 \approx 1/q
-(q-2)^qz_1/[q(q-1)^{q-1}]$, the critical line for $\rho_1 \ll
1$ shows a linear behavior $\rho_1 \approx 1/q -\rho_2$. In the
high density limit, we find the following densities on the coexistence
line: for the fluid phase $\rho_1 \approx 1-1/z_1$ and $\rho_2 \approx
1/z_1^{q-1}$, so that for this phase we find $\rho_1 \approx
1-\rho_2^{1/(q-1)}$. Therefore, the line reaches the point $\rho_1=1$,
$\rho_2=0$ with infinite slope. For the solid phase, we get
$\rho_1=(\rho_{1A}+\rho_{1B})/2 \approx 1/(2z_1)$ and $\rho_2 \approx
1/2-1/(2z_1)$, so that $\rho_1=1/2-\rho_2$. Qualitatively, these
features of the phase diagram are similar to the ones in Fig. 3 of the
paper by Poland \cite{p84}.

In general, the fixed point or limit cycle may be obtained solving
equation (\ref{fpeq1}) for the variable $g$. For a given value of
$z_1$, at small values of $z_2$ we find a single root for the equation
in the interval $[0,1]$. For values of $z_1$ below the tricritical
value, above the critical value of $z_2$, two roots are found, one of
them being a double root. They correspond to the fixed
point variables $x$ and $y$ at sublattices A and B. If $z_1>z_{1,TC}$,
for $z_2$ above the limit of stability of the solid phase and below
the limit of stability of the fluid phase, five roots
are found, such that the smaller and the larger ones correspond to
the fixed points associated to the solid phase, while the
intermediate root corresponds to the fluid phase. Between the
extremal and the central roots, there are two additional roots, which
are unstable. Finally, in the region where only the solid phase is
stable, three roots are found, the intermediate one being
unstable. These findings lead to an alternative procedure to
locate the tricritical point, since there the polynomial $h(g)$, its
first and second derivatives vanish. This is discussed in some more
detail in the appendix.

\begin{figure}
\begin{center}
\includegraphics[width=8.0cm]{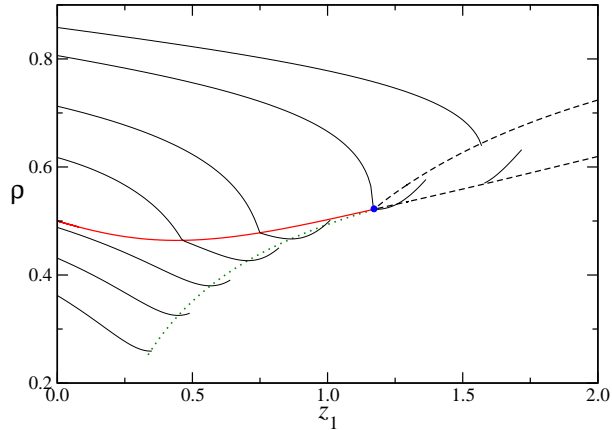}
\caption{(Color on line) Total density of particles $\rho=\rho_1+2\rho_2$ 
as a function of $z_1$ for fixed values of the pressure $\Pi$ (thin 
full lines) . The starting
points of the isobars corresponds to large particles only and at the endpoints
only small particles are present. The full line (red) corresponds
to the critical condition and the two dashed lines are the densities of the
coexisting phases. For isobars which cross the coexistence line, the minimum
is located on this line. The loci of the minima of the isobars is represented
by the dotted line (green). From lower to higher starting densities,
the isobars correspond to reduced pressures $\Pi=0.3,\,0.4,\,0.5,\,0.6,\,
0.7,\,0.86062$ (the isobar with the minimum at the tricritical point) and 
$1.0$. Bethe lattice with $q=4$.} 
\label{f6}
\end{center}
\end{figure}

Another interesting feature of the Bethe lattice solution of this model 
is that the isobaric curves of the total density of particles as a function 
of one of the fugacities show a non-monotonical behavior in the fluid phase. 
Let us define the total density of particles as $\rho=\rho_1+2 \rho_2$, 
so that it will be in the interval $[0,1]$. If we recall that the 
grand-canonical free energy per site is related to the pressure by:  
\begin{equation}
\phi_b=-\frac{pv_0}{k_BT},
\end{equation}
where $v_0=V/N$ is the volume per site, so that we may define $\Pi=-\phi_b$ 
to be the reduced pressure. For a fixed value of the pressure, 
we may now obtain the density of particles $\rho$ as a function of the 
activity $z_1$. Some resulting curves may be seen in figure \ref{f6}. We 
notice that the curves are not monotonic, starting with a negative slope 
at $z_1=0$. The minima in
the isobars are located on the dotted lines in the phase diagrams of 
figures \ref{f3} and \ref{f6}. We notice that this line starts at a 
particular point of the $z_1$ axis and ends at the tricritical point. 
Some aspects of these minima may be discussed analytically. To find 
the point of the $z_2=0$ axis where the minimum of the isobars is located, 
we may obtain a solution of the model for 
$z_2 \ll 1$, since we can solve it exactly for $z_2=0$. The fixed point
values of the ratios are, up to linear terms in $z_2$:
\begin{subequations}
\begin{eqnarray}
R_1 &\approx& z_1\left( 1-\frac{q-1}{(1+z_1)^q}z_2 \right), \\
R_2 &\approx& \frac{z_2}{(1+z_1)^{q-1}}.
\end{eqnarray}
\end{subequations}
Using these approximate solutions and the Eqs. (\ref{rho}) for the
densities, we may then find an approximate expression for the total density 
of particles:
\begin{equation}
\rho=\rho_1+2\rho_2 \approx \frac{z_1}{1+z_1}-\frac{(q-1) z_1-2}{(1+z_1)^{(q+2)}}
z_2.
\end{equation}
The minimum of the isobars correspond to the condition:
\begin{equation}
\left( \frac{\partial \rho}{\partial z_1}\right)_{\Pi}=0.
\end{equation}
This derivative may be calculated at $z_2=0$ noticing that:
\begin{equation}
\left( \frac{\partial \rho}{\partial z_1}\right)_{\Pi}=
\left( \frac{\partial \rho}{\partial z_1}\right)_{z_2}+
\left( \frac{\partial \rho}{\partial z_2}\right)_{z_1}
\left( \frac{\partial z_2}{\partial z_1}\right)_{\Pi},
\end{equation}
and since $d \phi_b=-\rho_1 dz_1/z_1-\rho_2 dz_2/z_2$, the last derivative 
in the expression above is:
\begin{equation}
\left( \frac{\partial z_2}{\partial z_1}\right)_{\Pi}=
-\frac{\rho_1z_2}{\rho_2z_1}.
\label{dz2dz1}
\end{equation}
Finally, we get:
\begin{equation}
\left( \frac{\partial \rho}{\partial z_1}\right)_{\phi_b}=
\frac{(q-1)z_1-1}{(1+z_1)^2},
\end{equation}
so that in the limit $z_2 \to 0$ the minimum of the isobars is located at 
$z_1=1/(q-1)$. A similar analysis may be done close to $z_1=0$, showing
that the slope of the isobars is negative there. As can be seen in 
Eq. (\ref{dz2dz1}), for constant pressure $z_2$ is a decreasing function 
of $z_1$, a feature which is valid in the solid phase also. Therefore,
any isobar which starts at $(z_1=0,z_2)$, will end at 
$(z_1,z_2=0)$ such that $\Pi(0,z_2)=\Pi(z_1,0)$. This is apparent in 
figure \ref{f6}, where we notice that each isobar for finite pressures 
ends at a finite values of $z_1$ and $\rho$. Isobars which start in the 
solid phase, at $(0,z_2)$ with $z_2>(q-1)^{q-1}/(q-2)^q$, will cross 
either the critical line or the coexistence line before they end in
the fluid phase. In the first case we notice a discontinuity in the slope 
of the isobars. In the second case, as expected the density is discontinuous 
as the coexistence line is crossed, so that the minimum is located on the 
coexistence line. 

It is worth mentioning that such non-monotonic behavior of the density
of particles for constant pressure as a function of a fieldlike variable
is found in nature, one of the most studied examples of this kind is
the density anomaly of liquid water, where a maximum is found in the isobaric
curves of density as a function of the temperature close to the 
freezing point\cite{d03}. 
In many studies in the literature, simple models were proposed which show
such anomalies, and it is believed that an interparticle potential with
two length scales may be the source of the thermodynamic anomalies. A
recent work of this kind, where also many earlier studies are referenced,  
may be found in \cite{ssob10}. For such models, the solution on tree-like
lattices may also be useful \cite{osb10}, and in this particular example
both maxima and minima of the isobaric curves of density as a function of
temperature were obtained. Although of course the present model is very
different from the lattice gases related to water, it is interesting
that here also two length scales are present in the interparticle
interactions.

\section{Conclusion}
\label{concl}
The Bethe lattice solution of the athermal lattice gas with two kinds
of particle, a small one occupying a single site and the other a site
and its 
first neighbors, shows a phase transition between a fluid phase, at
low values of the activity of the large particles $z_2$ and a solid
phase, which appears at higher values of $z_2$ and where one of the
sublattices is preferentially occupied by the large particles. The
results of the Bethe approximation for the particular case where no
small particles are present ($z_1=0$) are recovered \cite{b60}. The
transition remains continuous for small values of $z_1$, the critical
line starts with a negative slope as $z_1$ is increased, passes
through a minimum and ends at a tricritical point, so that for $z_1$
larger than the tricritical value the transition is discontinuous. 

The behavior of the densities of particles, as shown in figure
\ref{f5}, may be compared with similar results obtained 
using series expansions for the model defined on the square lattice,
shown in figure 3 in reference \cite{p84} by Poland. Besides the
expected quantitative differences, we notice in general a qualitative
agreement of both diagrams. A significant difference is that in our
calculation the lines corresponding to the fluid and the solid phases
meet at the tricritical point forming an angle, while in the diagram
by Poland a smooth junction is suggested. Since the dotted lines in
Poland's diagram are the result of an extrapolation, it seems that
this aspect may be worth to be studied in more detail using other
techniques. However, it may be possible that the angle we find here is a
characteristic of the mean field approximation implicit in Bethe
lattice calculations.

\section*{Acknowledgments}
We thank Prof. Ronald Dickman for calling our attention to
the model studied here, for helpful discussions, and a critical reading
of the manuscript. JFS is grateful for
partial financial support by the brazilian agency CNPq. 

\appendix
\section{Determination of the tricritical point}
\label{dtcp}
As mentioned in the text, one way to determine the tricritical
point is to solve the set of three equations for the polynomial
$h(g,z_1,z_2)$ defined in Eq. (\ref{fpeq1}):
\begin{subequations}
\begin{eqnarray}
h&=&0, \\
\frac{\partial h}{\partial g}&=&0, \\
\frac{\partial^2 h}{\partial g^2}&=&0.
\end{eqnarray}
\end{subequations}
Although it seems to be a rather simple task to solve this system of
nonlinear algebraic equations for $g$, $z_1$, and $z_2$, standard
numerical methods, based on Newton-Raphson procedures, often do
not converge to the expected (physical) solution. This may be due to 
the fact that the
first and second derivatives of $h$ with respect to $g$ vanish at the
solution. Therefore, we used another procedure, taking advantage of
the fact that $h$ is a polynomial in the variable $g$:
\begin{equation}
h(g,z_1,z_2)=\sum_{i=0}^N h_i(z_1,z_2)g^i,
\end{equation}
where $N=1+(q-1)^2[1+(q-1)^2]$. Now $g_0$, the value of $g$ at the
tricritical point, is a triple root of the polynomial, so that we may
write: 
\begin{equation}
h(g,z_1,z_2)=\sum_{i=0}^N h_i(z_1,z_2)g^i=
(g-g_0)^3\sum_{i=0}^{N-3}f_i(z_1,z_2)g^i
\end{equation}
Comparing terms with the same powers of $g$ in the equation above, we
may solve for the $N-2$ coefficients $f_i$ in terms of the coefficients
$h_i$ and $g_0$. The result is:
\begin{equation}
f_i(g_0,z_1,z_2)=-\sum_{j=0}^i c_{j-i}\frac{h_j}{g_0^{j-i+3}},
\end{equation}
for $i=0,1,2,\ldots,N-3$, and
\begin{equation}
c_j=\frac{(j+1)(j+2)}{2}.
\end{equation}
The remaining equations, corresponding to the powers $g^{N-2}$,
$g^{N-1}$, and $g^N$, are:
\begin{subequations}
\begin{eqnarray}
h_{N+2}-f_{N-5}-3g_0f_{N-4}-3g_0^2f_{N-3}&=&0 \\
h_{N-1}-f_{N-4}-3g_0f_{N-3}&=&0 \\
h_N-f_{N-3}&=&0.
\end{eqnarray}
\end{subequations}
The solution of these equations leads to the activities and the value
of the variable $g$ at the tricritical point.

\begin{table}
\begin{tabular}{ccccccccc}
\hline
\hline
$q$ & & $z_1$ & & $z_2$ & & $\rho_1$ & & $\rho_2$ \\
\hline
 3  & &  2.74598   & &  11.9231  & &  0.3937(8) & & 0.1476(5) \\
 4  & &  1.16956   & &  3.02938  & &  0.2985(7) & & 0.1117(5) \\
 5  & &  0.734355  & &  1.539662 & &  0.2407(3) & & 0.0894(3) \\
 6  & &  0.533384  & &  0.995756 & &  0.2013(3) & & 0.0746(3) \\
 7  & &  0.418259  & &  0.725526 & &  0.1729(2) & & 0.0641(3) \\
 8  & &  0.343861  & &  0.566972 & &  0.1515(2) & & 0.0561(2) \\
 9  & &  0.291885  & &  0.463718 & &  0.1349(2) & & 0.0499(2) \\
10  & &  0.253531  & &  0.391509 & &  0.1215(2) & & 0.0450(2) \\
\hline
\hline
\end{tabular}
\caption{Values of the activities and densities at the tricritical 
  point for Bethe
  lattices with different coordination numbers $q$. The error in  
  the activities values are $\pm1$ in the last decimal place.}
\label{t1}
\end{table}

\end{document}